\title[]{Software Rasterization of 2 Billion Points in Real Time}
\author[Markus Sch{\"u}tz \& Bernhard Kerbl \& Michael Wimmer]
{\parbox{\textwidth}{\centering Markus Sch{\"u}tz,
        Bernhard Kerbl,
        Michael Wimmer 
        }
        \\
{
    \parbox{\textwidth}{
        \centering TU Wien
    }
}
}

\begin{document}

% \teaser{
%  \vspace{-0.5cm}
%  \includegraphics[width=\textwidth]{images/teaser.jpg}
%         \caption{Software-rasterized point cloud. Each workgroup (128 threads) rasterizes blocks of 12.8k points. The size of the projected bounding box determines whether the workgroup utilizes full coordinate precision (green, 12 bytes) or lower coordinate precision (red, 4 bytes) for faster memory access. The middle and right images visualize the bounding boxes of the blocks/workgroups. TODO: better teaser}
%     \label{fig:teaser}
% }

\teaser{
 \vspace{-0.5cm}
 \includegraphics[width=\textwidth]{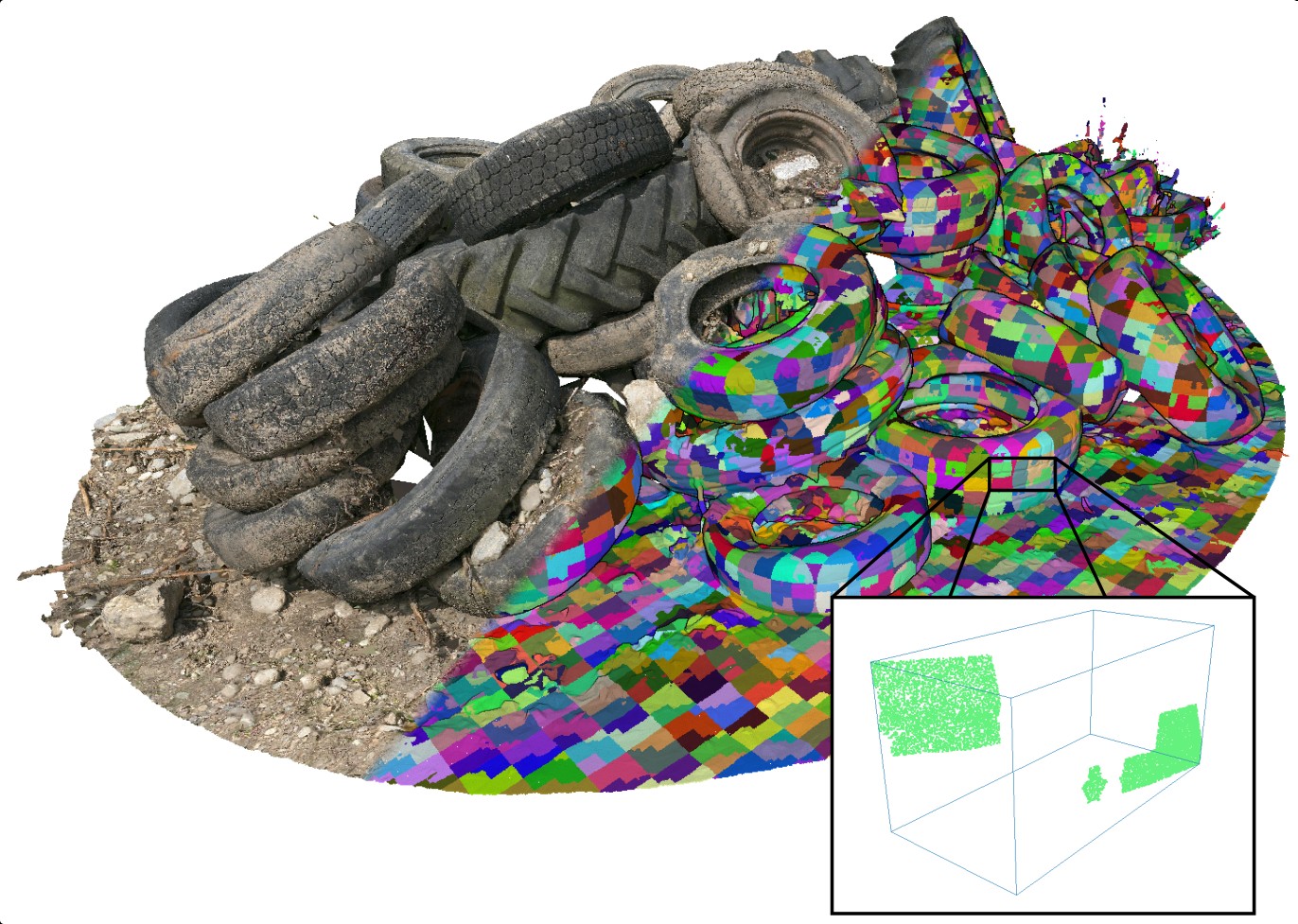}
        \caption{A point cloud grouped into batches of 10'240 points. Each batch is rendered by a single compute workgroup using 128 threads, and each thread renders 80 points for a total of 128 * 80 = 10'240 points per workgroup. Workgroups utilize the bounding box of their batches for frustum culling and to determine the required coordinate precision. Using 10 bit fixed-precision integer coordinates relative to the bounding box of a batch provides sufficient precision for the vast majority of visible batches. If 10 bits are insufficient, additional bits are loaded on demand. }
    \label{fig:teaser}
}

\maketitle
%-------------------------------------------------------------------------

\begin{abstract}

We propose a software rasterization pipeline for point clouds that is capable of brute-force rendering up to two billion points in real time (60fps). Improvements over the state of the art are achieved by batching points in a way that a number of batch-level optimizations can be computed before rasterizing the points within the same rendering pass. These optimizations include frustum culling, level-of-detail rendering, and choosing the appropriate coordinate precision for a given batch of points directly within a compute workgroup. Adaptive coordinate precision, in conjunction with visibility buffers, reduces the number of loaded bytes for the majority of points down to 4, thus making our approach several times faster than the bandwidth-limited state of the art. Furthermore, support for LOD rendering makes our software-rasterization approach suitable for rendering arbitrarily large point clouds, and to meet the increased performance demands of virtual reality rendering.

\begin{CCSXML}
<ccs2012>
<concept>
<concept_id>10010147.10010371.10010372.10010373</concept_id>
<concept_desc>Computing methodologies~Rasterization</concept_desc>
<concept_significance>500</concept_significance>
</concept>
</ccs2012>
\end{CCSXML}

\ccsdesc[500]{Computing methodologies~Rasterization}

\printccsdesc   
\end{abstract}

%\clearpage

\section{Introduction}

With the introduction of hardware with dedicated triangle rasterization units, hand-crafting rasterization routines in software became largely obsolete. Such custom-built rasterizers have nevertheless remained an ongoing topic of research in order to develop and study new rasterization approaches. Some of them eventually managed to beat hardware rasterization in specific scenarios~\cite{10.1145/1730804.1730817}, but in general, dedicated hardware remains the fastest approach. Nanite is the first framework that promises practical improvements for 3D games via hybrid software and hardware rasterization~\cite{karis2021nanite}. They found that directly rasterizing the fragments of a small triangle with atomic min-max operations can be faster than pushing the triangle through the hardware rendering pipeline. Therefore, only larger triangles are rasterized via hardware. 

Point-cloud models offer additional opportunities for efficient software rasterization, as the hardware rendering pipeline is largely dedicated to the rasterization of triangles and not points. In this paper, we consider point clouds as 3D models made of colored vertices, where each vertex is projected to exactly one pixel. Although this is a fairly strict limitation, it allows us to device algorithms that compete with graphics APIs that also only support one-pixel points, such as DirectX (POINTLIST primitive) and all backends that use it (WebGL, WebGPU, ANGLE, MS Windows games and applications, ...). We intent to support larger point-sprites in the future and use the evaluated performances of one-pixel points as a baseline for comparisons. Point clouds have no connectivity, so index buffers or vertex duplication are not required. The lack of a connected surface also makes uv-maps and textures irrelevant, which is why colors are typically directly stored on a per-vertex basis. Furthermore, point clouds acquired by laser scanners do not contain surface normals. Normals could be computed in a pre-processing step, but computed normals are not robust in sparsely sampled regions with high-frequency details such as vegetation, strings/cables/wires or even noise. We will therefore not consider normals in this paper, either.

Our approach builds on~\cite{SCHUETZ-2021-PCC} to further optimize several aspects of  software rasterization of points, which leads to an up to 3x higher brute-force performance. Specifically, our contributions to the state of the art of software rasterization of point clouds are:

\begin{itemize}
  \item Assigning larger workloads to batches to enable efficient batch-level optimizations.
  \item Adaptive coordinate precision coupled with visibility-buffer rendering for three times higher brute-force performance.
  \item Fine-grained frustum culling on batches of about 10'240 points, directly on the GPU. 
  \item Support for state-of-the-art level-of-detail structures for point clouds. 
\end{itemize}

\section{Related Work}

\subsection{Software Rasterization of Triangle Meshes}

Early CPU-side solutions for triangle rasterization in software were largely made obsolete in the 2000s by GPUs and their high-performance rasterization components.
However, their (partly hardwired) pipeline lacks the flexibility to perform custom hierarchical or context-dependent optimizations between individual stages.
The continuously advancing programmability of GPUs has given software rasterization its second wind:
\emph{Freepipe} demonstrated that for scenes containing many, small triangles, GPU software rasterization with one thread per triangle can outperform the hardware pipeline \cite{10.1145/1730804.1730817}.
CudaRaster and Piko expanded on this idea, introducing optimizations for hierarchical triangle rasterization, achieving competitive performance with hardware rasterization even for larger triangles \cite{10.1145/2018323.2018337, 10.1145/2766973}.
Complete software implementations of OpenGL-style streaming pipelines, including sort-middle binning and hierarchical rasterization, have been presented for NVIDIA CUDA and OpenCL \cite{Kenzel:2018, Kim2021}: rather than attempting to outperform hardware rasterization, these solutions aim to provide an environment for experimenting with extensions and optimizations that may be suitable for future hardware pipelines.
A comprehensive analysis of previous software rasterization approaches and the challenges they tackle is found in \cite{frolov20comparative}.

Most recently, software rasterization has received increased attention due to the launch of the Unreal Engine 5 and its virtual geometry feature, \emph{Nanite} \cite{karis2021nanite}. 
\emph{Nanite} provides both a hardware and a software pipeline for rasterization geometry and selects the proper route for rendered geometry dynamically.
In scenes with mostly pixel-sized triangles, their software pipeline reportedly achieves more than $3\times$ speedup. Its striking success begs the question whether high-performance software rasterization has not been overlooked as a viable method for other 3D representations as well.

\subsection{Software Raserization of Point Clouds}

Günther et al. proposed a GPGPU-based approach that renders points up to an order of magnitude faster than native OpenGL points primitives -- a feat that is possible because the fixed-function rendering pipeline is mainly targeted towards triangles~\cite{Gnther2013AGP}. Whenever a point modifies a pixel, their busy-loop approach locks that pixel, updates depth and color buffers, and then unlocks the pixel. Marrs et al. use atomic min/max to reproject a depth buffer to different views~\cite{Marrs2018Shadows}. Since only depth values are needed, 32 bit atomic operations are sufficient. Schütz et al. render colored point clouds via 64 bit atomic min-max operations by encoding depth and color values of each point into 64 bit integers, and using atomicMin to compute the points with the lowest depth value for each pixel inside an interleaved depth and color buffer~\cite{SCHUETZ-2021-PCC}. Our paper is based on this approach and the published source code\footnote{\url{https://github.com/m-schuetz/compute_rasterizer/releases/tag/compute_rasterizer_2021}}, and makes it several times faster for brute-forcing, but also adds support for frustum culling and LOD rendering. Rückert et al. also adapt this approach into a differentiable novel-view synthesis renderer that uses one-pixel point rendering to draw multiple resolutions of a point cloud, and neural networks to fill holes and refine the results~\cite{ruckert2021adop}. They also significantly reduce overdraw by stochastically discarding points whose assumed world-space size is considerably smaller than the pixel they are projected to. 

\subsection{Level-of-Detail for Point Clouds}

Rusinkiewicz and Levoy introduced QSplat, a point-based level-of-detail data structure, as a means to interactively render large meshes~\cite{QSplat}. They use a bounding-sphere hierarchy that is traversed until a sphere with a sufficiently small radius is encountered, which is then drawn on screen. Sequential Point Trees are a more GPU-friendly approach that sequentializes a hierachical point-based representation of the model into a non-hierarchical list of points, sorted by their level of detail. From a distance, only a small continuous subset representing a lower level of detail needs to be rendered, without the need for costly traversal through a dense hierarchical structure~\cite{Dachsbacher2003}. 

Layered point clouds~\cite{LPC} were one of the most impactful improvements to LOD rendering of point clouds, and variations of it still serve as today's state of the art. The original LPC constitutes a binary tree that splits the 3D space in half at each level of the hierachy. The key difference to the bounding-sphere hierarchy of QSPLATs is that each node itself is not a sphere, but a smaller point cloud comprising thousands of randomly selected points of the full point cloud. The large amount of geometry in each node reduces the amount of nodes that are required to store the full data set, which allows utilizing GPUs that are efficient at rendering hundreds of batches comprising thousands of points, each. Further research improved several aspects of layered point clouds, such as the utilized tree-structure, LOD generation times, and using density-based subsampling to properly support data sets without uniform density~\cite{Wand2008, Goswami2010, scheiblauer2011, Elseberg2013, MartinezRubi2015, Kang2019, Bormann:PCI}. Section~\ref{sec:lod} describes how our rasterization approach supports layered point clouds, using the \cite{Potree} structure as a specific example. 

\subsection{Coordinate Quantization}

Quantization is the conversion of a continuous signal to discrete samples. In case of coordinates, quantization typically refers to the conversion of floating point or double coordinates (the pseudo-continuous signal) to fixed-precision integer coordinates with a carefully chosen amount of bits and a uniform coordinate precision over the whole range. The uniform precision and control over the supported coordinate range, precision, and amount of bits makes quantization a frequently used part of coordinate compression schemes, complemented by delta and entropy encoding ~\cite{deering1995geometry, ISENBURG:LAZ} or hierarchical encoding~\cite{10.5555/581896.581904}. In this paper, we use quantization to encode coordinate bits such that they can be adaptively loaded, i.e., to load just a few bits if low precision is sufficient, or optionally load additional bits to refine the previously loaded low-precision coordinate. We do not, however, apply delta, entropy, hierarchical or similar encodings because they add additional computational overhead, and they make the decoding of one point dependant on the decoding of others, i.e., they can't be decoded individually.

\section{Method}

\begin{figure}
    \centering
    \begin{subfigure}[t]{\columnwidth}
        \includegraphics[width=\textwidth]{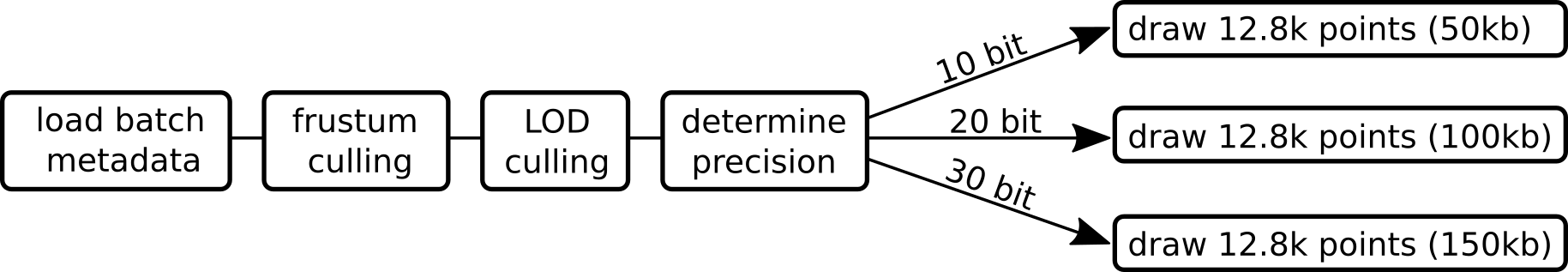}
        \caption{Flowchart of a single workgroup invocation.}
    \end{subfigure}
    \caption{Each workgroup renders one batch or octree node. If its projected bounding box is small, fewer coordinate bits are loaded, reducing memory bandwidth usage and boosting render performance accordingly.}
    \label{fig:pipeline}
\end{figure}

The core aspect of our rasterization method is that we assign larger batches of points to be rendered by each workgroup (e.g., 128 threads) instead of only one point per workgroup thread. These larger batches (e.g., 10k points -- 80 per thread) enable several optimizations that would be too costly on small batches (e.g., 128 points -- 1 per thread), but whose costs amortize with each additionally rendered point. Using larger batches with varying amounts of points also allows us to support several widely used level-of-detail structures, as discussed in Section~\ref{sec:lod}. Figure~\ref{fig:pipeline} gives an overview of the rendering steps within a workgroup. 

We will first describe our basic rendering pipeline, which we will then gradually expand by additional features and optimizations that ultimately allow us to render point clouds several times faster than the state of the art. 

\subsection{Data Structure}

\begin{figure}
    \centering
    \hfill
    \begin{subfigure}[t]{\columnwidth}
        \includegraphics[width=\textwidth]{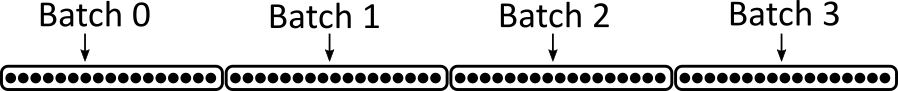}
        \caption{Unstructured Point Clouds}
    \end{subfigure}
    \par\bigskip
    \begin{subfigure}[t]{\columnwidth}
        \includegraphics[width=\textwidth]{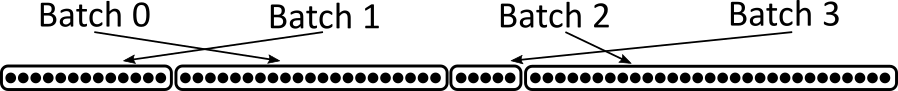}
        \caption{LOD Batches / Nodes}
    \end{subfigure}
    \caption{(a) For unstructured, Morton-code-ordered data sets, we group 10'240 consecutive points into a batch. (b) For LOD data, each octree node equals a batch and batches additionally store memory location and number of points inside the node/batch.}
    \label{fig:batches_memory}
\end{figure}

The data structure consists of a list of batches and a list of points. Each batch represents a number of consecutive points in the point list (see Figure~\ref{fig:batches_memory}), so for each batch we store the index of the first point of that batch, the number of points in that batch, and their bounding box. Each point consists of 4 attributes: low, medium, and high precision parts of the coordinates (details in Section~\ref{sec:vertex_compression}), and a color value. The 4 attributes are stored in a struct-of-arrays fashion, i.e., in separate buffers for low-precision part of the coordinate, another for medium precision part, etc., so that we only need to load components from memory during rendering that are actually needed. For the majority of points, this will just be the low-precision coordinate bits (4 bytes out of 16 bytes per point). 

For regular, unstructured point-cloud data sets, we suggest to simply group about 10'240 consecutive points into batches and compute their bounding box while loading the points. In practice, the majority of data sets we've encountered already provide sufficient locality, especially data sets generated through aerial LIDAR or photogrammetry. But not all of them do and almost all can benefit from sorting by Morton code (z-order) -- an easy to implement and efficient approach to create data sets with good locality~\cite{10.1145/588011.588037, 1532800, 10.1007/3-540-45033-5_3, LOM2020, SCHUETZ-2021-PCC}. For the remainder of this section, we will assume that data sets exhibit sufficient locality, either by default or through sorting by Morton-code, and we will further evaluate and discuss the impact of poor locality in Section~\ref{sec:evaluation}.

\begin{figure*}
    \centering
    \begin{subfigure}[t]{0.24\textwidth}
        \includegraphics[width=\textwidth]{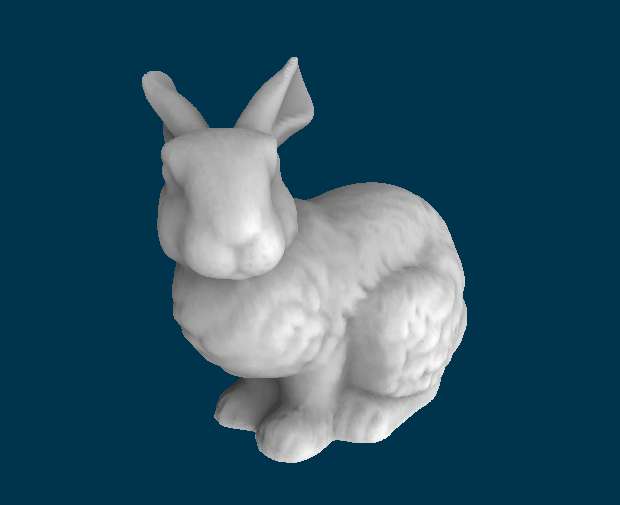}
        \caption{Point Cloud Data Set}
    \end{subfigure}
    \hfill
    \begin{subfigure}[t]{0.24\textwidth}
        \includegraphics[width=\textwidth]{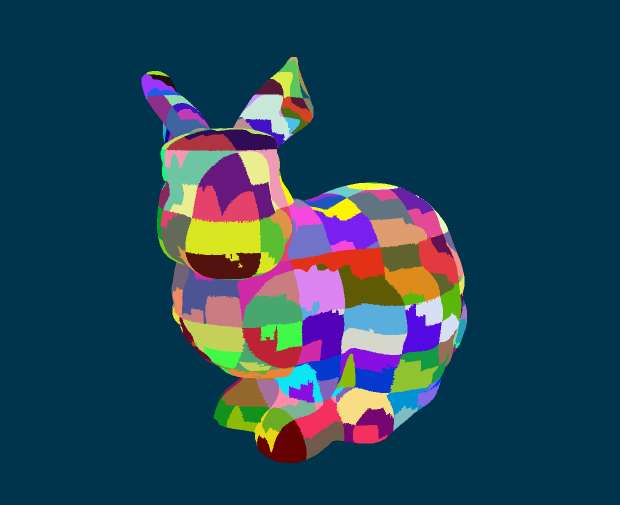}
        \caption{Batches}
    \end{subfigure}
    \hfill
    \begin{subfigure}[t]{0.24\textwidth}
        \includegraphics[width=\textwidth]{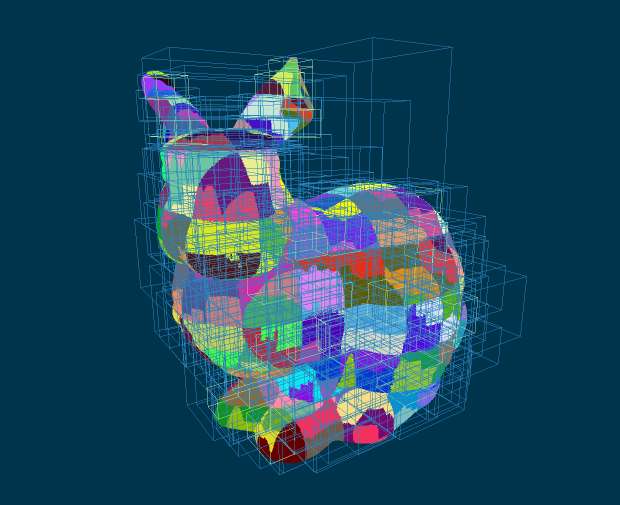}
        \caption{Bounding Boxes}
    \end{subfigure}
    \hfill
    \begin{subfigure}[t]{0.24\textwidth}
        \includegraphics[width=\textwidth]{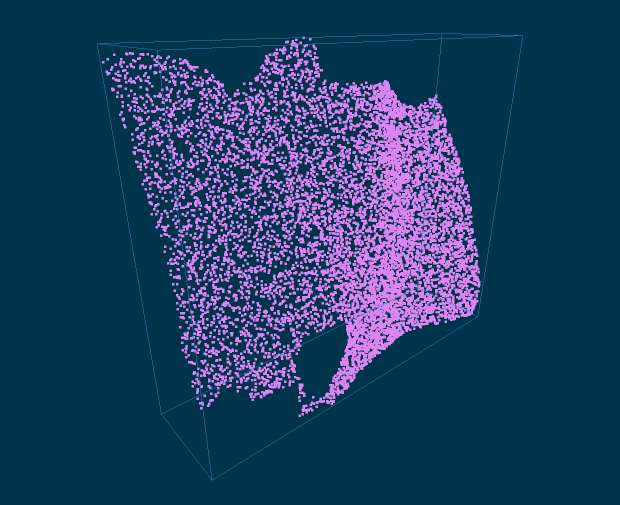}
        \caption{A single batch}
    \end{subfigure}
    \caption{A Morton-code ordered point cloud, grouped into batches of 10'240 consecutive points. The locality that is provided by the Morton order is sufficient for frustum culling and a bounding-box-based vertex compression/quantization scheme.  }
    \label{fig:batches_boxes_colors}
\end{figure*}

Figure~\ref{fig:teaser} and Figure~\ref{fig:batches_boxes_colors} show the results of grouping 10'240 consecutive points of a Morton-ordered point cloud into batches. Although the locality isn't perfect, the majority of batches are sufficiently compact with only a couple of outliers that experience extremely large jumps between different clusters of points. The main advantages of this approach are that it's trivial to implement, requires no preprocessing, and can be done while a data set is loaded with negligible impact on performance. 

\subsection{Basic Rendering Pipeline}
\label{sec:basic}

The basic pipeline spawns one compute workgroup for each batch to render its points. Each workgroup comprises 128 threads and each thread renders $n$ points. In practice, we found $n$ between 60 to 200 to perform equally well on an RTX 3090, and will therefore assume $n = 80$ points per thread (128 * 80 = 10'240 points per batch) throughout the paper. The rasterization process of each individual point is the same as in previous work \cite{SCHUETZ-2021-PCC}: Each point is projected to pixel coordinates, its depth and color value are encoded into a single 64 bit integer, and atomicMin is used to compute the point with the smallest depth value for each pixel, as shown in Listing \ref{lst:basicRasterization}. An early-depth test (a non-atomic depth comparison before an expensive atomic depth-compare and write) ensures that the majority of occluded points are rejected before invoking an expensive atomic update. 

\begin{lstlisting}[language=Java,label={lst:basicRasterization},caption={Point rasterization, including an early-depth test.},captionpos=b]
vec4 pos = worldViewProj * position;

int pixelID = toPixelID(pos);
int64_t depth = floatBitsToInt(pos.w);
int64_t newPoint = (depth << 32) | pointIndex;
// fetch previously written point
uint64_t oldPoint = framebuffer[pixelID];

// Early-depth test
if(newPoint < oldPoint){
    atomicMin(framebuffer[pixelID], newPoint);
}
\end{lstlisting}

The first improvement over \cite{SCHUETZ-2021-PCC} is that the larger workloads per workgroup allow implementing efficient workgroup-wise frustum culling with a granularity of 10'240 points. At the start of the workgroup invocation, we first load the bounding box of the current batch and abort right away if it does not intersect with the view frustum. Another improvement is that spawning fewer workgroups and giving each of them larger tasks reduces the scheduling overhead of the GPU.

\subsection{Adaptive Vertex Precision}
\label{sec:vertex_compression}

One of the main bottlenecks in previous work \cite{SCHUETZ-2021-PCC} is memory bandwidth usage. They reported a peak performance of 50 billion points per second using 16 bytes per point, which utilizes 85\% of the GPU's memory bandwidth. Since this speed is limited by bandwidth, any significant improvement of brute-force rendering performance requires some form of vertex compression.

\begin{figure}
    \centering
    \begin{subfigure}[t]{0.19\columnwidth}
        \includegraphics[width=\textwidth]{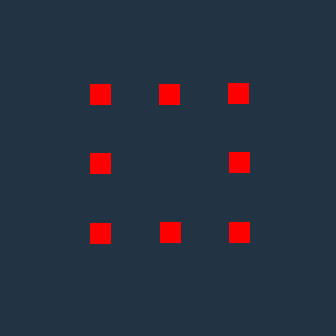}
        \caption{2 bits}
    \end{subfigure}
    \hfill
    \begin{subfigure}[t]{0.19\columnwidth}
        \includegraphics[width=\textwidth]{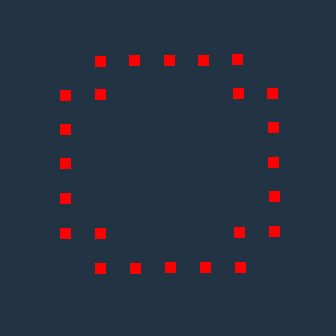}
        \caption{3 bits}
    \end{subfigure}
    \hfill
    \begin{subfigure}[t]{0.19\columnwidth}
        \includegraphics[width=\textwidth]{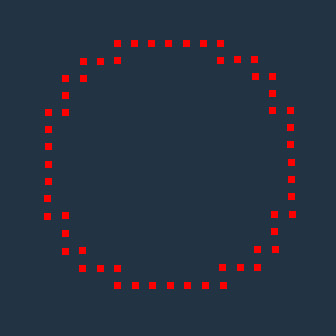}
        \caption{4 bits}
    \end{subfigure}
    \hfill
    \begin{subfigure}[t]{0.19\columnwidth}
        \includegraphics[width=\textwidth]{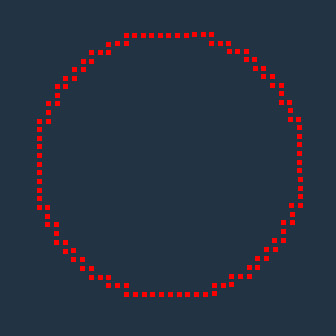}
        \caption{5 bits}
    \end{subfigure}
    \hfill
    \begin{subfigure}[t]{0.19\columnwidth}
        \includegraphics[width=\textwidth]{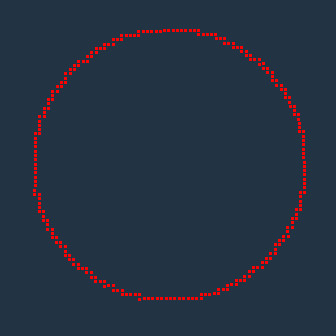}
        \caption{6 bits}
    \end{subfigure}
    \caption{Coordinate quantization provides a trade-off between precision and storage size. The amount of possible coordinate values per axis is given by $2^{bits}$.}
    \label{fig:quantization}
\end{figure}

\begin{figure}
    \centering
    \hfill
    \begin{subfigure}[t]{\columnwidth}
        \includegraphics[width=\textwidth]{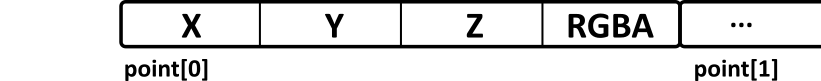}
        \caption{Common point-cloud memory layout with 16 bytes per point.}
    \end{subfigure}
    \hfill
    \par\bigskip
    \begin{subfigure}[t]{\columnwidth}
        \includegraphics[width=\textwidth]{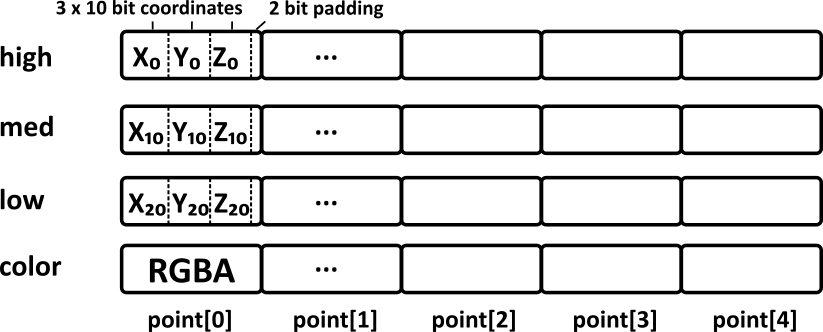}
        \caption{Splitting coordinate bits into a low, medium and high-precision buffer. Each buffer stores 10 bits per coordinate axis, encoded into 4 bytes per point. }
    \end{subfigure}
    \caption{(a) Point cloud renderers typically load at least 16 bytes per point during rendering. (b) Splitting coordinate bits into separate buffers allows loading just the required bits, depending on the viewpoint. 10-bit coordinates typically suffice for the majority of points. }
    \label{fig:memory_layout}
\end{figure}

We propose an adaptive precision scheme that allows us to load as many bits for coordinates as necessary for a given viewpoint. We achieve this by splitting the bits of a coordinate into three separate buffers storing the low, medium, and high precision parts, as shown in Figure~\ref{fig:memory_layout}. The low precision part always needs to be loaded. The medium precision part can be optionally loaded to recover some of the bits that we removed, and the high precision part is used to optionally recover the remaining bits. The different precision levels are established via coordinate quantization, i.e., by converting coordinates to fixed-precision integers with specific amounts of bits, and then splitting the quantized bits. By quantizing the point coordinates with respect to the bounding box of the batch instead of the bounding box of the whole 3D model, we can achieve a high coordinate precision with few bits. Figure~\ref{fig:quantization} illustrates that the resolution of the coordinates quickly approaches the resolution of the pixel grid. In our case, we first convert the coordinates within a batch to 30 bit fixed-precision integer coordinates that indicate each point's position within its batch. The X-coordinate is computed as follows, for example:

\begin{equation}
X_{30} = min( \lfloor 2^{30} * \frac{x - boxMin.x}{boxSize.x} \rfloor , 2^{30} - 1)
\label{eq:targetIndexPermute}
\end{equation}

These 30 bits are then split into three 10-bit components representing low, medium and high precision parts. The 10 most significant bits (bit indices 20 to 29) of this 30 bit integer are the low-precision bits. They tell us the coordinate of a point within a batch with a precision of $\frac{1}{2^{10}} = \frac{1}{1024}$ of its size. Considering that the majority of rendered batches in any given viewpoint are typically smaller than a couple of hundreds of pixels, and that 10 bits grant us 1024 different coordinate values, we find that 10 bits per axis are sufficient to render most points with sub-pixel coordinate precision. Due to buffer alignment recommendations, simplicity, and because the smallest accessible primitive values on GPUs are typically 32 bits, we then pack the 10-bit x, y and z components into a single 32 bit integer with the remaining 2 bits used as padding, as shown in Listing~\ref{lst:encoding}. The result is a 32-bit integer buffer where each 4-byte element contains the 10 lowest precision bits of the three coordinate axes of a single point.

\lstset{
  mathescape,         
  literate={->}{$\rightarrow$}{2}
           {ε}{$\varepsilon$}{1}
}

\begin{lstlisting}[language=Java,label={lst:encoding},caption={Encoding the most significant 10 bits of each axis into a single 32 bit integer. },captionpos=b]
uint32_t X$_{low}$ = (X$_{30}$ >> 20) & 1023;
uint32_t Y$_{low}$ = (Y$_{30}$ >> 20) & 1023;
uint32_t Z$_{low}$ = (Z$_{30}$ >> 20) & 1023;
uint32_t encoded = X$_{low}$ | (Y$_{low}$ << 10) | (Z$_{low}$ << 20)
\end{lstlisting}

Likewise, we generate two more buffers for the medium (bits 10 to 19, middle) and high precision bits (bits 0 to 9, least-significant). During rendering, each workgroup can now choose to load a single 4-byte integer containing just the low precision bits that grants 1'024 coordinate values, 2 such integers (low+medium precision) granting 1'048'576 coordinate values, or 3 such integers (low+medium+high precision), which theoretically allows for one billion different coordinate values. However, we should note that the usefulness of 30 bit precision is limited in practice because we still convert the integer coordinates to floating-point coordinates during rendering, where the conversion from a 30-bit fixed-precision integer to a 32-bit floating point value will incur a loss of precision due to rounding errors and the non-uniform distribution of precision in floats (high precision for small values, low precision for large values). This issue is not specific to our approach, however, as point clouds are typically already stored in integer coordinates on disk and the conversion to floats for rendering has always been an issue for point clouds with a large extent. Handling this issue is out of scope of this paper, but we would like to note that storing coordinates as 30 bit fixed-precision integers would allow us to convert them to highly accurate double-precision coordinates (e.g., for measurements), while traditional floating point storage causes an irrecoverable loss of precision. 

The coordinate precision during rendering is determined through the projected size of a batch. If a batch projects to less than 500 pixels, we use 10 bit coordinates. The reason for choosing 500 pixels as the threshold even though 10 bits can represent 1024 different coordinates values is that quantization changes the distance between any two points by up to the size of a quantization grid cell, i.e., points that were one grid cell's size apart might now be twice as far apart. Using half the size of the quantization grid as the pixel threshold ensures that we do not introduce additional holes between rasterized points.

Compared to previous work~\cite{SCHUETZ-2021-PCC}, this approach reduces the required memory bandwidth for the majority of rendered points from 16 bytes down to 8, since only 4 bytes instead of 12 are needed for most coordinates, and another 4 for colors. However, we can further cut this into half by using a visibility-buffer (item-buffer) approach~\cite{Burns2013Visibility, 10.1145/357332.357335}, i.e., we render point indices rather than colors into the framebuffer during the geometry pass, and transform the point indices to vertex colors in a post-processing step. Doing so reduces the amount of memory accesses to color values down from the number of processed points to the number of visible points.

\subsection{Vertex Pre-Fetching}
\label{sec:prefetch}

To identify performance bottlenecks of the above approach, we performed a direct port of its GLSL shader code to NVIDIA CUDA. 
Doing so enabled the use of the proprietary Nsight Compute tool to pinpoint suboptimal behavior in our routines.
Our investigation into dominant stall reasons revealed that performance is---unsurprisingly---governed by memory operations, with approximately $70\%$ of the total kernel run time spent on them.

Since we choose the batch size to be a multiple of the work group size, threads persist and process multiple points in a loop before returning. 
Naively, the corresponding point data is fetched and rasterized to the framebuffer in each iteration.
Batches simply define a linear range in memory, thus their contents can be transferred by workgroups with perfect coalescence.
Furthermore, the compression scheme outlined in Section \ref{sec:vertex_compression} economizes on available bandwidth when fetching point data. 
However, loading each processed point individually at the start of its associated iteration still incurs a direct dependency of the steps in Listing \ref{lst:basicRasterization} on global device memory latency.
Due to the early-depth test in its body and the parameterizable iteration count, the compiler cannot trivially unroll the loop without creating secondary issues associated with complex program flow (e.g., frequent instruction cache misses).
Hence, we perform manual pre-fetching of point data. For the lowest precision (using 32 bits per point), we execute a vectorized 128-bit load in each thread, yielding enough data for four successive iterations; for medium-precision points, a 128-bit load fetches the data for two iterations.
This policy significantly alleviates reported stalls due to memory latency and simultaneously reduces the number of memory instructions, without compromising on coalesced accesses.
Overall, we found pre-fetching to cause $\approx 30\%$ performance gain.
A welcome side effect, though less impactful, is the reduction of updates to the frame buffer by this policy, if points are spatially sorted. Since threads now load and process consecutive points in memory, the early-depth test has a higher chance of failing: threads are now more likely to query (and find) information in the L1 cache for pixels that they wrote to in a previous iteration.

We note that, with pre-fetching in place, the main remaining bottleneck is the code block for early-depth testing and framebuffer updates. Although these accesses are somewhat localized if points are spatially ordered, a residual irregularity remains in their access pattern.
The coarse-grained memory transfer policy of the GPU consequently causes more than $2\times$ the amount of actually accessed information to be transferred.
However, this update pattern is inherent to the overall routine design, and any further attempts to improve on it without extensive revision led to diminishing returns.

\subsection{Level-of-Detail Rendering}
\label{sec:lod}

In this section we will describe how support for some of the most popular LOD structures for point clouds -- Layered Point Clouds (LPC)~\cite{LPC} and its variations -- can be added to our point rasterization approach. Layered point clouds are a hierarchical, spatial acceleration structure where points with varying density are stored in the nodes of the tree. Lower levels contain coarse low-density subsets of the whole point cloud. With each level, the size of the nodes decreases, leading to an increase of the the density of points as the number of points in each node stays roughly the same. The structures are often additive -- meaning that higher LODs increase detail by rendering more points in addition to lower LODs instead of replacing them -- but replacing schemes are also possible. 

We implement and evaluate our support for LPC structures based on the Potree format, which constitutes a variation of LPC that uses an octree and populates nodes with subsamples of the point cloud with a specific, level-dependant minimum distance. Each octree node comprises about 1k to 50k points, and Potree itself typically renders about 100 to 1000 nodes for any given viewpoint. Up until now we have assumed that each batch renders exactly 10'240 points, but we can easily adapt our rasterization approach to support the Potree structure by allowing varying amounts of points per batch, as shown in Figure~\ref{fig:batches_memory}. In addition to the bounding box, we will also need to store the number of points as well as the memory location of the first point for each batch. The workgroup size of 128 threads remains the same, but each thread will now render $\lceil \frac{batch.numPoints}{128} \rceil$ points instead of exactly 80. 

The Potree format is structured such that octree nodes whose projected bounding boxes are small can be entirely ignored, because the points stored in their parents will already provide sufficient detail. This means that the procedure in Section~\ref{sec:vertex_compression} that is used to determine the coordinate precision during rendering based on the screen-size of the bounding box can now be used to entirely cull the node. We suggest to cull the nodes with following conditions in mind: Each Potree node is cubic and stores a subsample of points with a resolution that approximately matches a grid with $128^3$ cells, and our rasterizer maps each point to exactly 1 pixel. To avoid holes between rendered points, we therefore suggest to cull those nodes that are smaller than 100 pixels on screen. However, it is also viable to cull larger octree nodes on lower end GPUs to improve performance and cover up the resulting holes in a post-processing pass, e.g., via depth-aware dilation or more sophisticated hole-filling approaches~\cite{ruckert2021adop, AutoSplats, Rosenthal2008, grossman1998point, Pintus2011, Marroquim:2007:pbg}.

\subsection{High-Quality Shading}

LOD rendering works well in conjunction with high-quality splatting (point-sprites or surfels~\cite{Botsch:HQS}) or shading (one-pixel points~\cite{SCHUETZ-2021-PCC}), a form of color filtering for point clouds that blends overlapping points together. Colors and amount of points inside a pixel within a certain depth range (e.g., 1\% behind the closest point) are summed up during the geometry processing stage, and a post-processing shader then divides the sum of colors by the counters to compute the average. Schütz et al. \cite{SCHUETZ-2021-PCC} suggest two variations: one that uses two 64-bit atomicAdd instructions per point into four 32 bit integers to sum up color values and counters, and another variation that uses a single 64 bit atomic instruction per point to compute the sum of up to 255 points, with a fallback that uses two 64-bit atomics if more than 255 points contribute to the average. However, when using an LOD structure, the amount of overlapping points with similar depth is essentially guaranteed to be lower than 255, so we can safely use high-quality shading with just a single 64-bit atomic add instruction per point. This limit can even be raised to up to 1023 points by using the ``non-robust" variation without overflow protection. For unstructured data sets, we suggest to use the variation with overflow protection due to the potentially large amount of overlaps.

\subsection{Virtual Reality Rendering}

Virtual reality (stereo) rendering greatly increases the performance requirements -- even more so for point clouds as they typically suffer from aliasing artifacts that need to be addressed to provide an acceptable VR experience. In addition to a specific configuration of our pipeline (use LOD rendering for performance, high-quality shading to reduce aliasing and large framebuffers for additional anti-aliasing via supersampling), we can exploit specific properties of VR rendering in our approach.

First, the scene needs to be drawn twice -- once for each eye. Instead of duplicating the entire rasterization pipeline by calling the compute shaders twice, we suggest to modify the shader to simply draw each point into both framebuffers. While this doesn't double the performance, it provides a significant improvement, as discussed in Section~\ref{sec:evaluation}. 

Second, in a VR setup, details in the periphery aren't as noticeable as details in the center of the view. This is partially because most details are perceived in the gaze direction, i.e., mostly straight ahead in VR, but also because the rendered image will be distorted before it is shown inside the HMD to counter the lens distortion~\cite{Vlachos:2015:AVR}. The applied distortion compresses peripheral regions, thus reducing the amount of detail in the image. We therefore suggest to render peripheral regions of the framebuffer with reduced geometric complexity by raising the threshold for LOD culling, e.g., culling nodes smaller than 300 pixels in the periphery, nodes smaller than 100 pixels in the center of the view, and 200 pixels in between. The resulting holes between points are then filled in post-processing, in our case via a simple depth-aware dilation shader that increases point sizes to up to 7x7 in the periphery and 3x3 in the center. Depth-aware means that the closest point within the pixel ranges are expanded.

\section{Evaluation}
\label{sec:evaluation}

\begin{figure*}
    \centering
    \begin{subfigure}[t]{0.19\textwidth}
        \includegraphics[width=\textwidth]{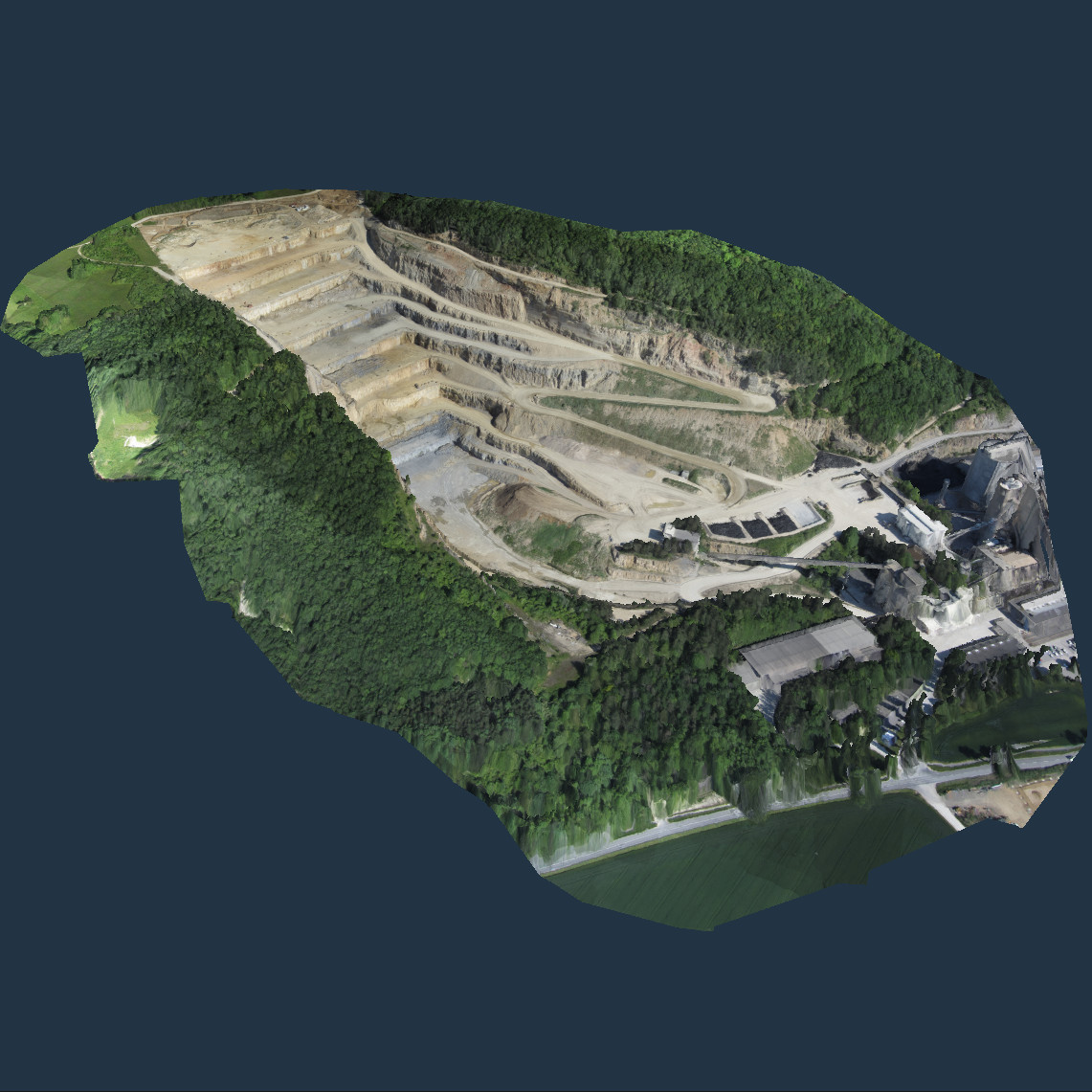}
        \caption{Eclepens}
    \end{subfigure}
    \hfill
    \begin{subfigure}[t]{0.19\textwidth}
        \includegraphics[width=\textwidth]{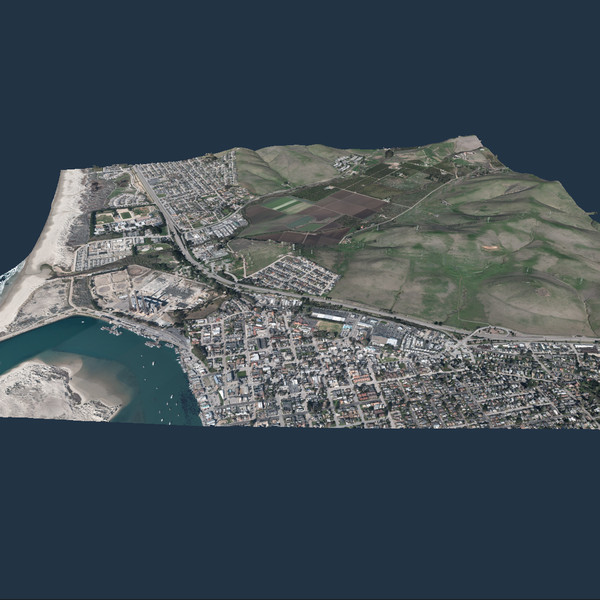}
        \caption{Morro Bay }
    \end{subfigure}
    \hfill
    \begin{subfigure}[t]{0.19\textwidth}
        \includegraphics[width=\textwidth]{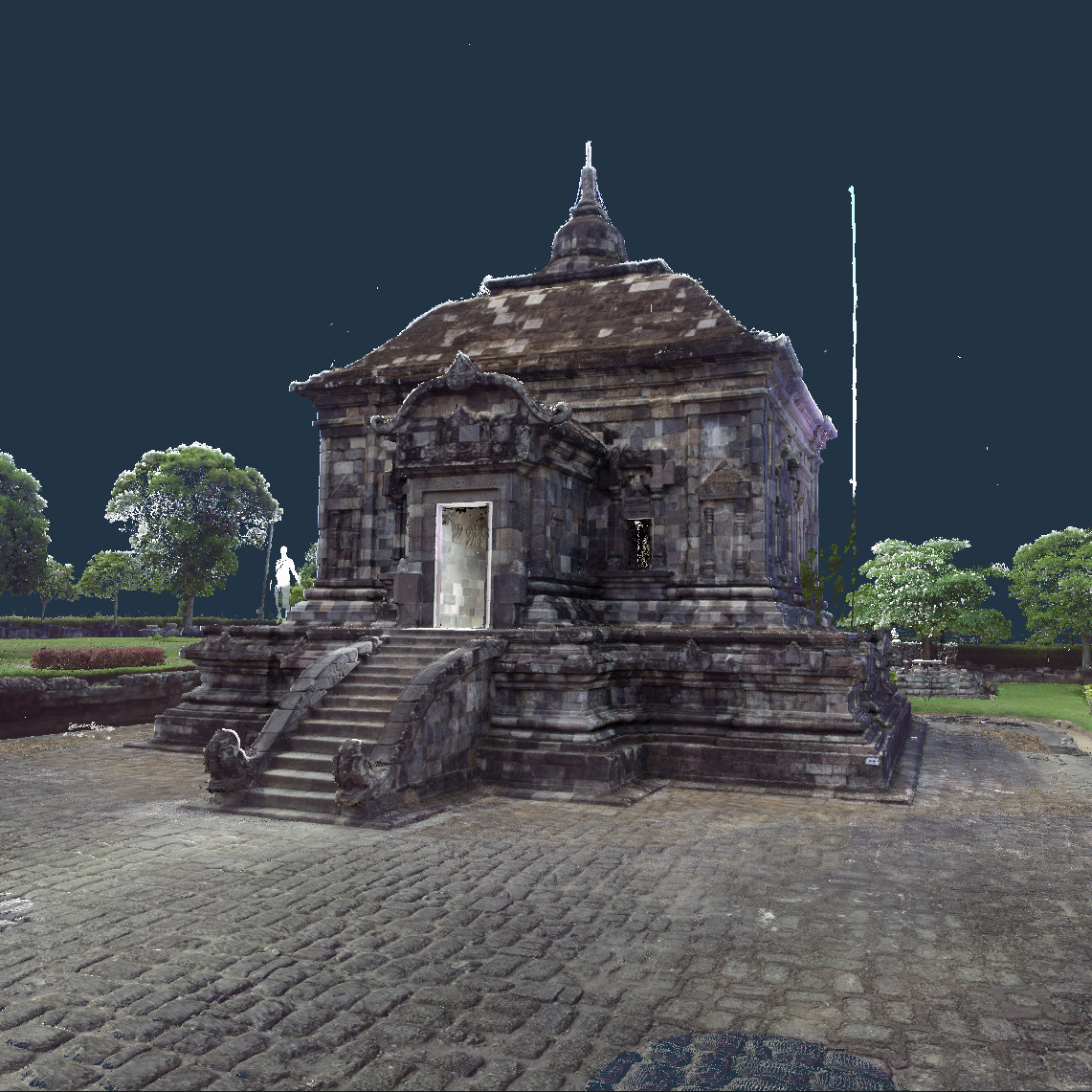}
        \caption{Banyunibo outside}
    \end{subfigure}
    \hfill
    \begin{subfigure}[t]{0.19\textwidth}
        \includegraphics[width=\textwidth]{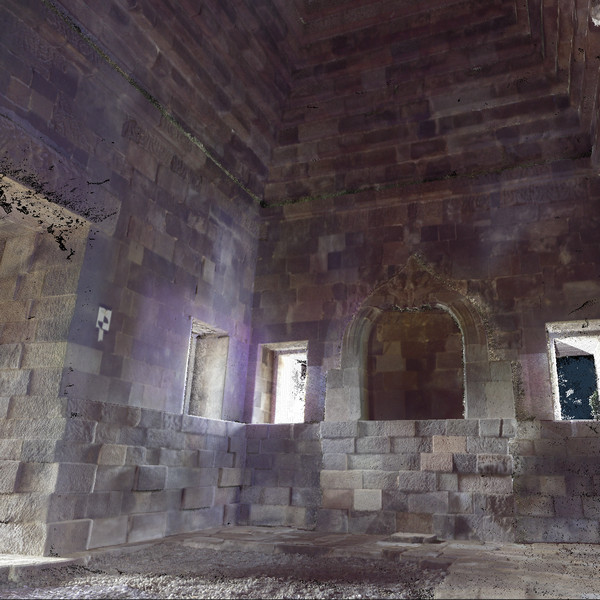}
        \caption{Banyunibo inside}
    \end{subfigure}
    \hfill
    \begin{subfigure}[t]{0.19\textwidth}
        \includegraphics[width=\textwidth]{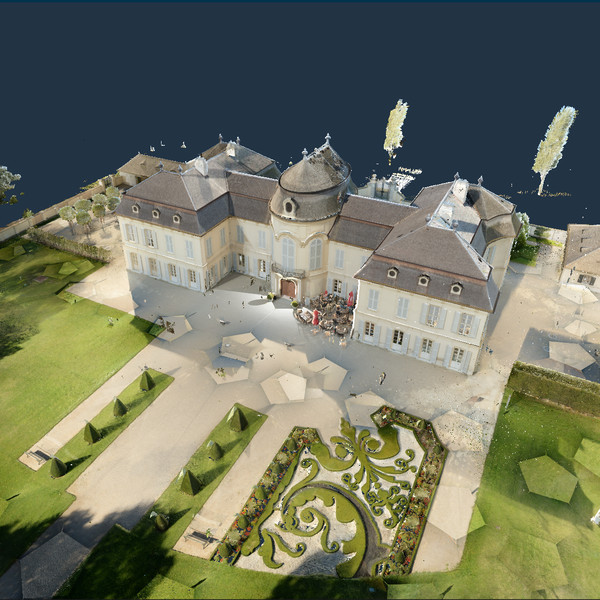}
        \caption{Niederweiden}
    \end{subfigure}
    \hfill
    \caption{Test data sets. (a) A quary captured with photogrammetry. (b) A coastal city, captured with aerial LIDAR. (c+d) A candi in indonesia, captured with photogrammetry and a terrestrial laser scanner. (e) A manor captured with terrestrial lasser scanning. }
    \label{fig:test_data_sets}
\end{figure*}

The proposed method was evaluated with the test data sets shown in Figure~\ref{fig:test_data_sets}. The smaller data sets, Eclepens and Morro Bay, are relatively easy to handle due to their low depth complexity, i.e., the number of hidden surfaces is typically small. Niederweiden poses a bigger challenge due to the higher point density and an interior room that is either occluded when viewed from the outside, or it occludes everything else when viewed from the inside, but occluded points are still processed due to the lack of occlusion culling. 

The performance was computed through OpenGL timestamp queries at the start and end of each frame. All durations represent the average time of all frames over the course of one second. The evaluation was conducted on the following test systems:

\begin{itemize}
    \item NVIDIA RTX 3090 24GB, AMD Ryzen 7 2700X (8 cores), 32GB RAM, running Microsoft Windows 10.
    \item NVIDIA GTX 1060 3GB, AMD Ryzen 5 1600X (6 cores), 32GB RAM, running Microsoft Windows 10.
\end{itemize}

Unless specified otherwise, all reported timings are from the RTX 3090 system. The GTX 1060 (3GB) was only capable of keeping the smallest test data set, Eclepens (68M points), in memory. 

\subsection{Rasterization Performance}
\label{sec:raster_performance}

Table~\ref{tab:our_performance} shows the results of rendering various data sets with our proposed basic (with frustum culling and adaptive precision; Sections ~\ref{sec:basic},~\ref{sec:vertex_compression}), prefetch (basic + each thread prefetches 4 points at a time; Section~\ref{sec:prefetch}) and LOD (Section~\ref{sec:lod}) approaches, and compares it with a previous one~\cite{SCHUETZ-2021-PCC} and with GL\_POINTS. For unstructured point-cloud data, our approach with the prefetch optimization performs the fastest in all cases -- up to three times faster than previous work in overview scenarios, and five times faster for the inside viewpoint that benefits from frustum culling. 

Table~\ref{tab:performance_details} shows how many nodes and points were rendered in a frame from the given viewpoint. Processed nodes include all batches/nodes of the data set since we spawn one workgroup per node. Rendered nodes are those that pass the frustum and LOD culling steps. Frustum culling can reduce the amount of rendered nodes to slightly less than half for unstructured point clouds (Banyunibo inside), or down to several thousand out of hundreds of thousands of nodes in conjunction with LOD structures. The throughput is computed by taking the number of processed points in Table~\ref{tab:performance_details} and dividing it by the rendering time in Table~\ref{tab:our_performance}. On the RTX 3090 system and with the prefetch method for unstructued data sets, we get throughputs of 69, 126.8, 144.7, 97.5 and 142.3 billion points per second for the five scenarios -- all of them larger than the peak throughput of 50 billion points per second reported in prior work~\cite{SCHUETZ-2021-PCC}. We can also look at the throughput in terms of how many points we are able to render in real-time (60fps) by mapping the results from points per second to points per $\frac{1000}{60} \approx 16$ milliseconds, which results in 1.1, 2.0, 2.3, 1.6, 2.3 billion points per 16 milliseconds. Thus, in three out of five scenarios we were able to render two billion points in real-time.

\begingroup
\begin{table*}
\begin{tabular*}{\textwidth}{@{\extracolsep{\fill}} |l|rr|l|r|rrr|rr|rr|}
\hline 
                     &  \multicolumn{2}{c|}{}  &              &             &  \multicolumn{3}{c|}{ours (unstructured)} & \multicolumn{2}{c|}{ours (LOD)}  & \multicolumn{2}{c|}{\cite{SCHUETZ-2021-PCC}}   \\
 Data Set            &  points   & size        &   GPU        &  GL\_POINTS &   basic     & prefetch           &  hqs   &    basic      &         hqs                  &          dedup & hqs       \\
 \hline                         
 Eclepens            &       69M &       1.1GB &    RTX 3090  &        34.9 &         1.1 &                1.0 &    2.8 &           0.7 &         1.4                  &            1.9 &       7.6 \\
 Eclepens            &       69M &       1.1GB &    GTX 1060  &        72.2 &         6.5 &                5.1 &   16.7 &           2.1 &         5.3                  &            9.5 &      26.2 \\
 Morro Bay           &      279M &       4.4GB &    RTX 3090  &       231.7 &         2.7 &                2.2 &    9.6 &           0.8 &         1.9                  &            6.0 &      33.5 \\
 Banyunibo (outside) &      529M &       8.5GB &    RTX 3090  &       198.3 &         4.2 &                3.3 &    9.0 &           1.3 &         3.3                  &           10.7 &      25.4 \\
 Banyunibo (inside)  &      529M &       8.5GB &    RTX 3090  &        67.9 &         2.6 &                2.2 &    6.1 &           2.1 &         4.6                  &           11.1 &      24.4 \\
 Niederweiden        &     1000M &        16GB &    RTX 3090  &       873.9 &         8.2 &                6.4 &   20.9 &           1.9 &         4.5                  &           19.8 &      69.1 \\
\hline 
\end{tabular*}
\caption{Comparing rendering times (in milliseconds) of GL\_POINTS with our new approach and the old one by \protect\cite{SCHUETZ-2021-PCC}. Framebuffer size: 2560 x 1140 (2.9MP). All point clouds sorted by Morton code.}
\label{tab:our_performance}
\end{table*}
\endgroup

\begingroup
\begin{table*}
\begin{tabular*}{\textwidth}{@{\extracolsep{\fill}} |l|r|r|r|r|r|r|r|r|}
\hline 
                     &  \multicolumn{4}{c|}{unstructured} & \multicolumn{4}{c|}{LOD}  \\
\hline 
                     &  \multicolumn{2}{c|}{nodes} & \multicolumn{2}{c|}{points} &  \multicolumn{2}{c|}{nodes} & \multicolumn{2}{c|}{points} \\
\hline 
 Data Set            &  processed   & rendered   &  processed  &  rendered &  processed   & rendered   &   processed  &  rendered  \\
\hline 
 Eclepens            &         6.7k &   6.7k    &       68.7M &     13.0M &       23.5k &       422  &         3.9M &       1.4M \\
 Morro Bay           &        27.2k &   27.2k   &      278.5M &     12.2M &       93.5k &       577  &         5.6M &       1.9M \\
 Banyunibo (outside) &        51.7k &   46.6k   &      477.6M &     13.7M &      195.7k &       3.3k &        26.9M &       4.4M \\
 Banyunibo (inside)  &        51.7k &   20.9k   &      214.5M &     19.1M &      195.7k &       8.2k &        67.1M &      12.5M \\
 Niederweiden        &        97.7k &   88.9k   &      910.7M &     51.6M &      346.6k &       2.1k &        27.3M &       6.4M \\
\hline 
\end{tabular*}
\caption{Detailed stats about the processed and rendered nodes and points during a frame. Processed nodes: All fixed-size batches (unstructured) or variable-sized octree nodes (LOD). Rendered Nodes: Nodes that passed frustum and LOD culling. Processed points: All points that were loaded. Rendered points: Points that pass point-wise frustum culling and early-depth, i.e., all points for which atomicMin is called. }
\label{tab:performance_details}
\end{table*}
\endgroup

\subsection{The Impact of Vertex Ordering}

\begin{figure}
    \centering
    % \begin{subfigure}[t]{\columnwidth}
    %     \includegraphics[width=\textwidth]{images/orderings/banyunibo.jpg}
    %     \caption{Eclepens}
    % \end{subfigure}
    % \hfill
    \begin{subfigure}[t]{0.49\columnwidth}
        \includegraphics[width=\textwidth]{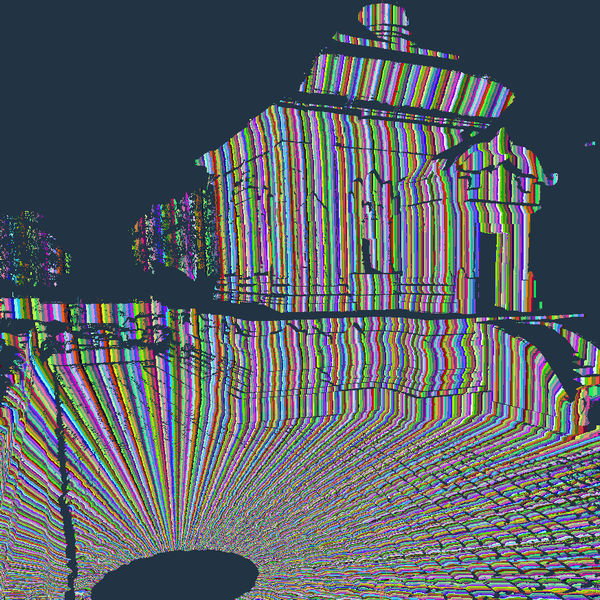}
        \caption{Batches - Scan Order}
    \end{subfigure}
    \hfill
    \begin{subfigure}[t]{0.49\columnwidth}
        \includegraphics[width=\textwidth]{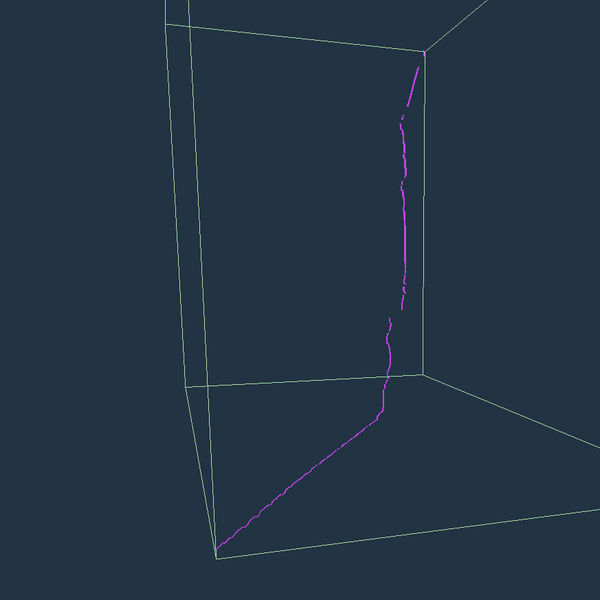}
        \caption{Batch - Scan Order}
    \end{subfigure}
        \begin{subfigure}[t]{0.49\columnwidth}
        \includegraphics[width=\textwidth]{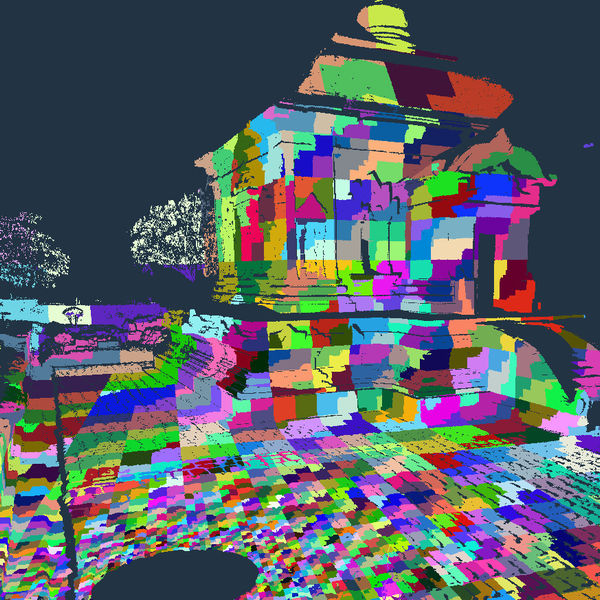}
        \caption{Batches - Morton Order}
    \end{subfigure}
    \hfill
    \begin{subfigure}[t]{0.49\columnwidth}
        \includegraphics[width=\textwidth]{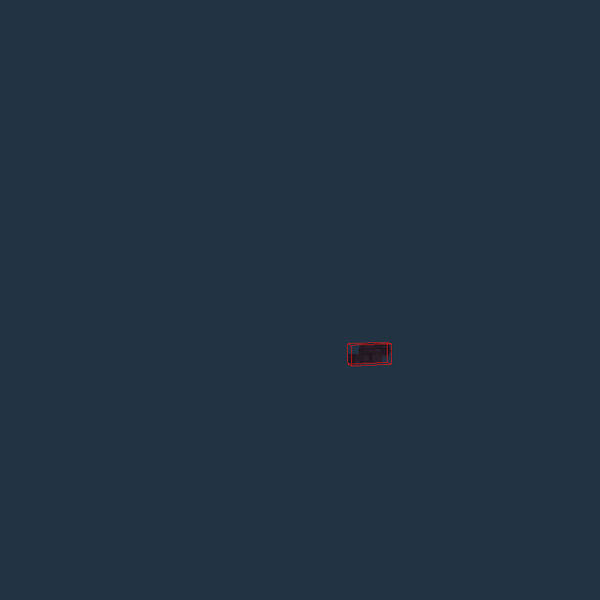}
        \caption{Batch - Morton Order}
    \end{subfigure}
    \caption{Terrestrial laser scanners typically scan along two rotation axes. (a+b) Poor locality in scan order results in huge but mostly empty batches. (c+d) Morton order efficiently establishes locality and results in compact batches. Both batches in (b) and (d) contain the same amount of points. }
    \label{fig:ordering}
\end{figure}

The disadvantage of our naïve approach of generating batches by grouping 10'240 consecutive points is that the resulting performances depend on the vertex order of the data set. Figure~\ref{fig:ordering} illustrates the vertex ordering of a terrestrial laser scanner that scans on two rotational axis, first top-bottom (pitch) and then right-left (yaw). Due to this, 10'240 consecutive points usually form a 3-dimensional curve along the surface of the scanned object. The resulting batch has a large extent with mostly empty space. The next batch is formed by the next "scan-line", with an almost identical bounding box. The bottom row of Figure~\ref{fig:ordering} demonstrates vertex order and the resulting batches after the points are sorted by Morton order. The resulting batches are more compact with less empty space, and therefore more suitable to frustum culling and rendering with lower coordinate precision.

We evaluated the performance differences between scan-order and Morton-order on a subset of the Banyunibo data set comprising only of the laser scans. From an outside viewpoint, scan-order requires 5.5ms to render a frame and Morton-order requires 3.9ms, which is mostly attributed to the lower coordinate precision requirements of the compact batches. From an inside viewpoint, the scan-order requires 7.8ms to render a frame and Morton-order requires 2.1ms. In this case, the even greater performance differences can further be attributed to frustum culling, which culls about 68\% of all nodes of the Morton-ordered data set, but only 35\% nodes of the scan-ordered data set.

\subsection{VR Performance}

VR rendering requires several times higher performance since we need to render data sets with higher frame rates, twice per frame, and in high quality. We evaluated the VR performance of our approach on a Valve Index HMD (1440×1600 pixels per eye) with an RTX 3090 GPU. The targeted framerate is 90fps, i.e., each frame needs to be finished in 11.1 milliseconds, or closer to about 9ms to account for additional computations and overhead in the VR pipeline. The targeted resolution is 2478 x 2740 (6.8MP) per eye, mandated by the 150\% resolution setting in SteamVR. The high resolution alleviates some aliasing issues, but high-quality shading is also needed and used to avoid severe flickering artifacts.

We evaluated the VR performance with an outside-in view of the Candi Banyunibo data set, comprising 529 million points in 195k octree nodes. 21.4 million points in 2k nodes passed the frustum and LOD culling routines. After early-depth testing, 8 million points were drawn with atomicMin. The total frame time was 8.3 milliseconds, which provides sufficient reserves for additional computations and overhead of the VR rendering pipeline. Rendering the depth buffer for the HQS shader took 1.7 ms for both eyes, or 1.3 ms when rendering just a single eye, which demonstrates the benefits of drawing each point to both render targets within a single compute shader invocation. Similarly, drawing the color buffer of the HQS approach took 2.5 ms for both eyes and 1.5 ms for a single eye. The resolve pass, which enlarges the more sparsely rendered points in the periphery and stores the colors into an OpenGL texture, takes about 2.7 ms per frame for VR rendering, which is mainly attributed to the large and dynamic dilation kernels of 3 x 3 (center) to 7 x 7 (periphery) pixels. Finally, clearing the relatively large depth and color buffers (2468 x 2740 per eye) takes about 0.6 ms per frame.

\section{Discussion}

In this section we'd like to discuss several issues and limitations, as well as potential improvements that we have not evaluated, yet.

First, we believe that this is a significant improvement for point cloud rendering, but it's not useful for games in its current state. Point clouds require a large amount of colored vertices to represent geometry without holes and sufficient color-detail, while meshes can use a single textured triangle in place of thousands of points. However, massive amounts of point cloud data sets exist as a result of scanning the real world, and this paper provides tools to render them faster without the need to process the data into LOD structures or meshes. Still, we hope that the presented approach might provide useful insights in future developments of hybrid hardware+software rasterization approaches for triangle meshes. For example, we could imagine that an approach like adaptive coordinate precision could be used to treat smaller batches like a point cloud and only load data that is relevant for triangles for larger batches. 

The quality for virtual reality currently suffers from lack of proper color filtering. Although high-quality shading is applied and improves the results via blending, the issue is that the LOD structure removes most of the overlapping points, thus the blended result is not representative of all points, including the discarded ones. The results can be improved by applying color filtering (e.g. computing averages) to points in lower levels of detail during the construction of the LOD structure~\cite{QSplat,Wand2008,schuetz-2019-CLOD} (similar to mip mapping). Furthermore, implementing continuous LOD could improve the visual quality through a subtle transition in point density between LODs, and by eliminating popping artifacts as details are added and removed while navigating through the scene \cite{schuetz-2019-CLOD, LOM2020}.

The adaptive coordinate precision approach leads to significant performance improvements through a reduction in memory bandwidth usage, but it does not compress the points -- coordinates still use up 12 bytes of GPU memory. At this time, we deliberately did not employ sophisticated compression approaches such as delta and entropy encoding~\cite{deering1995geometry, ISENBURG:LAZ} or hierarchical encoding~\cite{10.5555/581896.581904} due to their additional computational overhead and because they make it impossible to decode coordinates individually. But in theory they could work. Delta and entropy encoding require to decode the points sequentially, which could work on a per-thread basis as each thread renders about 80 points sequentially. Generally, we think that compression could work on a per-batch (10'240 points) basis, a per thread (80 points) basis and/or a per-subgroup (32 or 64 cooperating threads) basis. 

% - better hole filling. it's pretty bad right now
% - evaluate actual compression vs. adaptive precision
% we've largely got higher coordinate precision than the loaded data (30 bit in batch vs 32 bit in whole model)

\section{Conclusions}

We have shown that software rasterization using OpenGL compute shaders is capable of rendering up to 144.7 billion points per second (Section~\ref{sec:raster_performance}), which translates to 2.3 billion points at 60 frames per second (16 ms per frame). The data structure is simple and generated on-the-fly during loading for unstructured point clouds, but LOD structures may also be generated in a preprocessing step for further performance improvements. Peak performances were observed in Morton ordered data sets, but many other orderings, for example aerial lidar scans that are sorted by timestamp and split into tiles, also provide substantial performance improvements by exploiting the spatial locality between consecutive points in memory. Data sets without sufficient locality (e.g., terrestrial laser scans) can simply be sorted by Morton order. 

The source code to this paper is available at~\url{https://github.com/m-schuetz/compute_rasterizer}.

\section{Acknowledgements}

The authors wish to thank \emph{Schloss Schönbrunn Kultur- und Betriebs GmbH, Schloss Niederweiden} and \emph{Riegl Laser Measurement Systems} for providing the data set of Schloss Niederweiden, the \emph{TU Wien, Institute of History of Art, Building Archaeology and Restoration} for the Candi Banyunibo data set~\cite{isprs-archives-XLII-2-W15-555-2019}, \emph{Open Topography} and \emph{PG\&E} for the Morro Bay (CA13) data set~\cite{CA13}, \emph{Pix4D} for the Eclepens quarry data set, Sketchfab user \emph{nedo} for the \emph{old tyres} data set (CC BY 4.0), and the\emph{Stanford University Computer Graphics Laboratory} for the \emph{Stanford Bunny} data set.

This research has been funded by the FFG project \textit{LargeClouds2BIM} and the Research Cluster “Smart Communities and Technologies (Smart CT)” at TU Wien.

\printbibliography

\end{document}